**V.V. BEZGUBA [1,2,*] and O.A. KORDYUK [1,2,\*\*]**

[1] Kyiv Academic University,
  36 Academician Vernadsky Blvd., UA-03142 Kyiv, Ukraine
[2] G.V. Kurdyumov Institute for Metal Physics of the N.A.S. of Ukraine,
  36 Academician Vernadsky Blvd., UA-03142 Kyiv, Ukraine

* werevv@gmail.com, ** kordyuk@kau.edu.ua, kordyuk@imp.kiev.ua

# MULTIBAND QUANTUM MATERIALS

Quantum materials are defined by the emergence of new properties resulting from collective quantum effects and by holding promise for their quantum applications. Novel superconductors, from high-$T_c$ cuprates and iron-based superconductors to twisted monolayers, exhibit a higher level of emergent complexity, with a multiband electronic structure playing a pivotal role in their comprehension and potential applications. Here, we provide a brief overview of key multiband effects in these superconductors and topological semimetals, offering guidelines for the theory-assisted development of new quantum materials and devices.

## 1. Introduction

Although the foundation for understanding the properties of all materials over the past hundred years has been the quantum theory of condensed matter, the term 'quantum materials' has only recently come into use [1]. This is a collective term that encompasses superconductors, strongly correlated electronic systems, materials with electronic ordering, topological insulators (all of which can be referred to as 'electronic quantum materials'), as well as other systems like ultracold atoms. What unites these materials is the concept of 'emergence' [2] — the emergence of new properties resulting from collective quantum effects.

On the other hand, this term has evidently emerged under the influence of the popularity and rapid development of 'quantum technologies', often referred to as the 'second quantum revolution' [3], primarily encompassing quantum computing and communication. Therefore, in the definition of quantum materials, the concept of emergence is often reinforced by the promise of their quantum applications.

Classical superconductivity is the most known emergent phenomenon that emerges from quantum pairing of electrons (fermionic quasiparticles) by phonons (the acoustic waves' quanta) into pairs that behave as bosons. So, while classical superconductors are the first example of quantum materials that come to mind, the novel superconductors: high-Tc cuprates (Cu-SC) [4] or iron-based pnictides and chalcogenides (Fe-SC) [5], exhibit the next level of emergent complexity, where 'multibanding', the multiple-band electronic structure, is important for both pairing mechanisms [6] and quantum applications [7].

The potential of superconducting quantum computer implementations is evident as major computer companies, including Google [8] and IBM [9], develop their quantum computers using classical superconductors, particularly aluminium. However, they face decoherence issues [10, 11] that are less likely to be resolved with classical superconductors. In this context, novel multiband superconductors show significant promise.

Here, we provide a review of multiband effects in novel quantum materials, such as multiband superconductors and topological semimetals, in order to gain a deeper understanding of the underlying physical mechanisms behind their emergent properties and for the development of future quantum applications.

## 2. Emergence of 'Multibanding'

In their normal state, the materials under consideration are metals. 'Multiband metals' can be defined as materials with multiple bands crossing the Fermi level, which translates to metals with a multisheet Fermi surface.

While most theoretical models deal with an idealized single-sheet Fermi surface, multisheet Fermi surfaces are not uncommon. Among pure metals, only a few exhibits a single-sheet Fermi surface, including all alkali metals and some transition metals (e.g., the Cu group) [12]. In others, multiple Fermi surface sheets appear as consequence of its interaction with the Brillouin zone boundaries or when several bands of similar binding energy cross the Fermi level.

Figure 1 summarizes additional reasons for emergence of multibanding in single band crystals that we will consider here. In simplest cases, two conduction bands or two Fermi surfaces may arise due to the doubling of the unit cell [13], bilayer splitting [14], or band inversion [15]. In general, the presence of any superstructure can lead to the emergence of small Fermi surface pockets that approach the structure to topological Lifshitz transitions [16]. Bilayer splitting can be generalized to bands with the same orbital origin and similar energy, like Ta $5d_{xy}$ and $5d_{x2-y2}$ bands in $2H$-TaSe$_2$ [17] or Fe $3d_{xz}$ and $3d_{yz}$ bands in the iron-based superconductors [5]. The band inversion can be extended to general overlapping of bands of different origin.

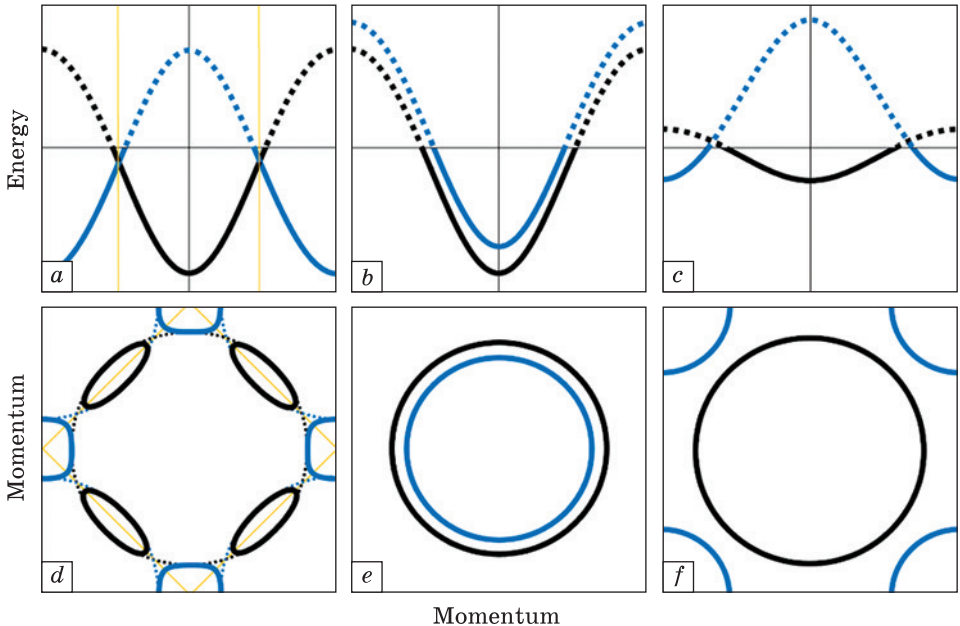

*Fig. 1.* Emergence of 'multibanding' in metals; two conduction bands (*a*–*c*) or two Fermi surfaces (*d*–*f*) may arise due to the doubling of the unit cell (*a*, *d* — yellow lines show the boundary of the new Brillouin zone), bilayer splitting (*b*, *e*), and a band inversion (*c*, *f*); for distinction of the colour in this and subsequent figures, the reader is referred to the web version of this article

Naturally, having multiple bands does not necessarily lead to the emergence of new physics. To achieve that, these bands should possess properties that are critically different with respect to key interactions, such as symmetry, orbital character, effective mass/density of states, and topology. Here, we provide a brief overview of multiband structures that can be significant for both pairing mechanisms and quantum applications. This includes superstructured crystals with electronic phase competition, superconductors with different order parameters, and bands with tuneable topologies.

## 3. Multiband Materials and Effects

### 3.1. Superstructured Superconductors

Transition metal dichalcogenides (TMDs) are classical quasi-$2D$ compounds that exhibit a diverse range of superstructured or electronically ordered phases driven by multiband electronic structure [17, 18]. The close proximity of superconductivity (SC) and charge density wave (CDW) on the electronic phase diagrams of many of TMDs ($2H$-NbSe$_2$ [13], $1T$-TaS$_2$ [19, 20], $1T$-TiSe$_2$ [21], $1T$-Ta(Se,Te)$_2$ [22]) is generally considered indicative of the competition between these electronic orders for the phase space. However, the mechanisms behind the emergence or enhancement of superconductivity by CDW have been proposed many times [13].

Interestingly, for both Cu-SC and Fe-SC, the situation is quite similar. Just instead of coexisting with CDW, the SC phase typically neighbours the spin density wave (SDW) phase [5, 6], as shown in Fig. 2, raising a common question: 'Is the electronic spatial ordering a friend or a foe of superconductivity?' [23].

Bilayer twisted graphene [24] serves as a recent example of an engineered multiband material. Its superstructure emerges from the Moiré pattern created by the interaction between tilted monolayers. The emergence of superconductivity at 'magic angles' in twisted graphene is attributed to the formation of flat bands at the Fermi level within the superstructure-defined Brillouin zone [25], providing the case when superstructure can be a friend for superconductivity.

With the growing interest in twisted graphene, TMDs have also become a hot topic as potential Moiré systems [26–28]. The buzz around TMDs stems from their unique combination of favourable electronic and mechanical properties, making them both intriguing subjects for fundamental research and valuable components in advanced electronics and optoelectronics, thanks to their atomic-scale thickness and promising properties. Recent experiments on heterobilayer TMD Moiré systems reveal competition between Fermi fluid, chiral spin liquid, spin density wave, and charge density wave, and consider having potential as a solid-state-based quantum simulator [28].

In general, the competition between various phases poses an intriguing challenge for our understanding, while also offering promising opportunities for theory-assisted engineering of novel materials and electronic components that leverage their unique properties.

### 3.2. To Room Temperature Superconductor

While the aspiration to achieve room-temperature superconductivity has successfully pressed the limit to 250 K in lanthanum hydride [29], the classical phonon-mediated system, more then 35-years research on the initially more promising Cu-SC has yielded more modest results [4]. 134 K is the highest value of $T_c$ for cuprates reached in $HgBa_2Ca_2Cu_3O_{1+x}$ (with three $CuO_2$ layers per unit cell) at ambient pressure [30] that increases only up to 166 K at 23 GPa [31].

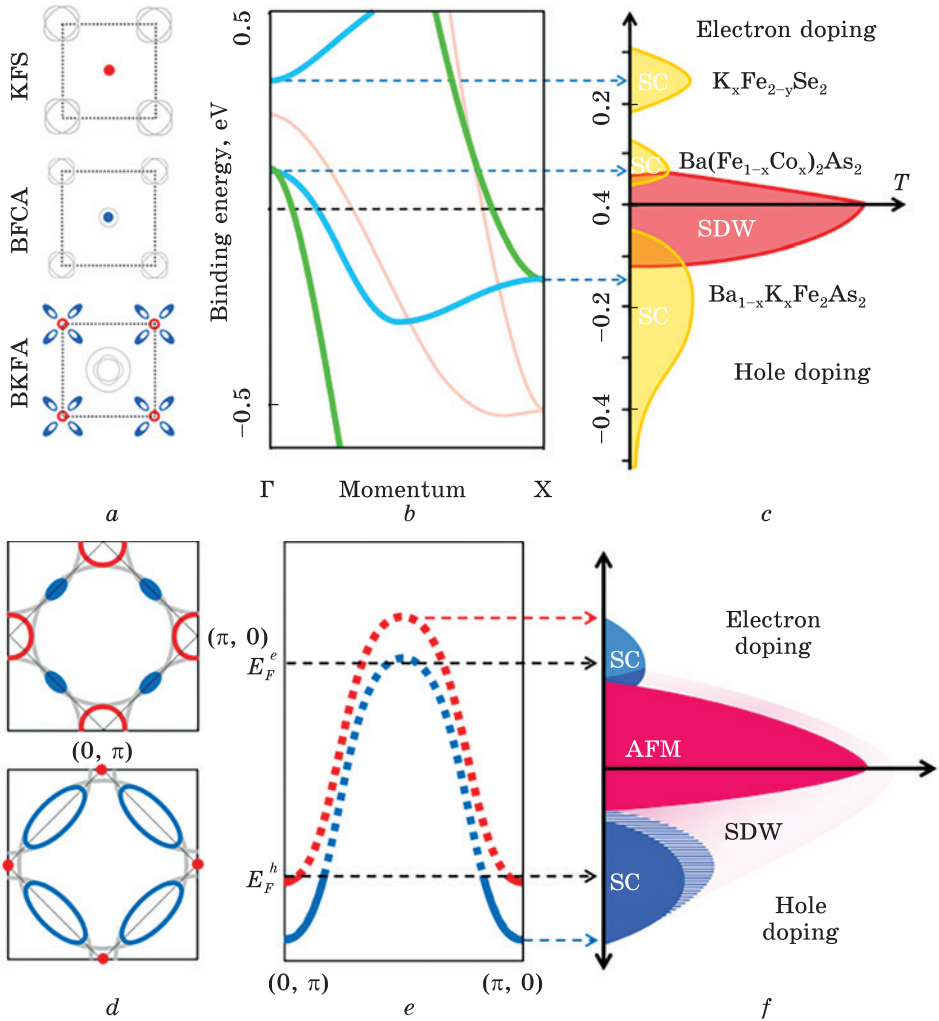

**Fig. 2.** (Colour online) The correlation of proximity to Lifshitz transition with $T_c$ is illustrated through a projection of the Fermi level crossing the 'rigid' electronic band structure of Fe-SC (*b*) and Cu-SC (*e*) on the charge carrier concentration scale of their phase diagrams (*c*) and (*f*); the corresponding 'optimal' Fermi surfaces (blue for hole- and red for electron-like sheets) are shown on (*a*) and (*d*), respectively (after Ref. [6])

Nonetheless, the extensive research on cuprates has uncovered a wealth of new insights into the complexity of electronic interactions in solids, providing valuable knowledge that we must comprehend and harness. In particular, the significance of multibanding became evident in the mid-1990s when the critical temperature displayed a notable non-monotonic dependence on the number of $CuO_2$ layers per unit cell. For instance, in the case of the mentioned mercury-based cuprates, $Hg-Ba_2Ca_{n-1}Cu_nO_{2n+2+\delta}$, at ambient pressure, the highest values of $T_c$ were 95 K, 127 K, 134 K, and 130 K for $n = 1, 2, 3,$ and 4, respectively [31]. This is evidently an outcome of the multilayer splitting, which cannot be solely attributed to a straightforward increase in the density of states with $n$.

In bilayer cuprates, the conducting band splits into bonding and antibonding counterparts with even and odd wave function symmetries. This splitting is in line with an increase in $T_c$ within the spin-fluctuations pairing mechanism [32]. However, despite experimental support for the spin-fluctuation scenario [33], some researchers questioned the adequacy of the electron coupling strength to the spin-fluctuations [34]. In addition, the SDW phase, if competing with superconductivity [23], would not help to increase the strength of this coupling.

The iron-based superconductors [35], which, among other expectations, are regarded as a long-sought solution to the high-$T_c$ problem [5], could finally provide insights into the significance of SDW-induced multibanding for cuprates [36].

The band structure of iron-based superconductors is notably more intricate than that of cuprates, typically consisting of five conduction bands crossing the Fermi level [5, 6]. However, this complexity provides an exceptionally fertile ground for establishing valuable empirical relationships. Notably, an empirical rule associates electronic structure with $T_c$, suggesting that the highest $T_c$ is achieved when the electronic structure is in proximity to a topological Lifshitz transition, as shown in Fig. 2. This can be also characterized by the presence of both large and small Fermi surfaces with distinct symmetries [5], while the enhancement of $T_c$ can be explained in terms of 'shape resonance' [37].

Currently, there is no Fe-SC compound that contradicts this empirical rule. Moreover, all Cu-FSs also seem to adhere to this rule, assuming that the normal state for superconductivity in cuprates is the SDW state. It's important to note that, while the SDW state on the electron side of the cuprates' phase diagram is readily observed in photoemission experiments [36], it remains somewhat elusive on the hole side but can be inferred through data analysis [38]. The broadening of spectra in the hole-doped SDW state may be attributed to dynamic antiferromagnetic correlations [39], which are not rapid enough to influence significantly the folding of the electronic structure or the pairing. To gain a clearer understanding, time-resolved ARPES experiments with instant Fermi surface mapping [40] could be instrumental in elucidating this scenario.

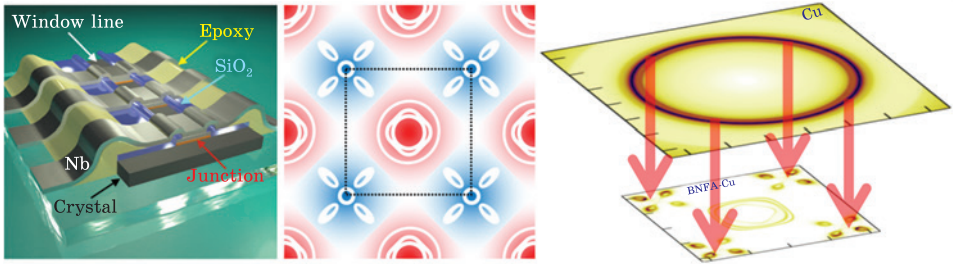

*Fig. 3.* (Left) Sketch of the BNFA−Cu/Nb JJ [7]. (Centre) The Fermi surface of BNFA (white lines) on top of the sign of the superconducting order parameter. (Right) Illustration of direct electron transitions governed by the effective Fermi surface of polycrystalline copper. A colour version of this figure can be viewed online

### 3.3. Multiple Order Parameter and Multiband Josephson Junctions

A distinctive characteristic of novel superconductors is a multiple and sign-changing superconducting order parameter. Its amplitude can be observed in ARPES spectra as superconducting gap, but its sign can be only explored by a phase sensitive probes like Josephson tunnelling [41]. The observation of this sign change in the order parameter and its distribution across the Fermi surface sheets provides evidence for magnetic pairing in both copper- and iron-based superconductors. This also holds significant implications for quantum electronic applications.

One potential solution of the decoherence issue in superconducting quantum computers involves using superconductors with larger superconducting gaps, which, compared to state-of-the-art aluminium technology, could greatly improve communication frequency between qubits, increasing their thermal fluctuation resistance. We have recently shown that it is possible to create a high-quality Josephson junction between Nb and a single crystalline iron-based superconductor $Ba_{1-x}Na_xFe_2As_2$ (BNFA) [41] and that its performance can be significantly adjusted by a Fermi-surface filter provided by a Cu intermediate layer (see Fig. 3) [7]. An exciting direction for this project is to replace the metallic layer with a semiconducting one, where a 2DEG forms a Fermi surface whose size can be easily controlled by gating.

Another potential solution to the issue of decoherence is the long-sought realization of Majorana fermions within a topological superconductor [42].

### 3.4. Topological Semimetals and Band Tuned Topology

Topological semimetals (TSMs) are a general name for all types of semimetals where energy bands cross while being topologically stable. Therefore, these are gapless systems which can vary significantly based on the band crossings degeneracy, nodes being represented by points or lines, etc. [43]. As a result, a lot of TSMs can be classified into distinct types with varying physics and electronic properties: Weyl and Dirac semimetals (different degeneracy of band crossings), nodal-line and nodal-surface semimetals (where Weyl and Dirac are nodal-point semimetals), type-I and type-II semimetals (slope of the band), and some more [15].

The first experimentally observed Weyl semimetal with inversion symmetry breaking and Fermi arcs was TaAs [44]. Fermi arcs are non-trivial surface states, which are essentially projections of connected chiral points such as Weyl points onto Fermi surface. Such surface states are one of the main points of interests towards topological materials, as topologically protected surface states arise from bulk and are extremely robust to external factors. Hereafter a lot of other Weyl semimetals studies started emerging such as TaP, NbAs, and NbP [45, 46].

Even before TaAs, three-dimensional Dirac semimetal phase was observed in $Cd_3As_2$ [47]. Later, it was shown that it is remarkably sensitive to the P doping [48]. BiTeI, which was previously identified as a polar semiconductor with giant bulk Rashba splitting emerged as a topological $3D$ Dirac semimetal hosting two, well isolated from each other and rest of the band structure, Dirac points [49]. Most well known Dirac semimetals include also $Na_3Bi$ [50], $ZrTe_5$ [51] and others.

All of the above are examples of type-I Dirac and Weyl semimetals, while type-II TSMs are possible as well (with a tilted Dirac/Weyl cone along **k**-direction). Lorentz invariant is violated in such case for quasi-particles near the Dirac/Weyl point. Such materials were predicted by [52] and soon after experimental observations followed for $WTe_2$ [53], $YbMnBi_2$ [54], $PtTe_2$ [55, 56], $PtBi_2$ [57].

One can see that TSMs come in great variety and complexity of their topologically non-trivial electronic structures. The same can be said about HTSCs and other multiband quantum materials. Such electronics structure complexity makes it an ideal platform for the search of desired material's properties.

Most iron-based superconductors have a gap between $p_z$ and $d_{xz}/d_{yz}$ bands [58], which turns out to be topological for them (for example, $FeTe_{0.5}Se_{0.5}$ [59]). Though the experiments do not confirm $FeTe_{0.5}Se_{0.5}$ carrying surface states [60], it might be possible to realize by doping the material and thus shifting inverted gap to the Fermi level. Similar methods can be tried on other IBSs to obtain potentially topological surface states and thus, topological superconductor with Majorana fermions.

Another example is $PtTe_2$. As shown in Ref. [56], Fermi level can be shifted substantially with Pt-doping and thus making possible the study of unique properties of type-II Dirac fermions. Fermi-level shifting should be possible to use for other transition-metal dichalcogenides too.

## 4. Conclusions

The novel quantum superconducting materials, ranging from high-temperature superconductors to twisted monolayers, exhibit a higher level of emergent complexity, with a multiband electronic structure playing a pivotal role in their understanding and potential applications.

The comparison between cuprates and iron-based superconductors leads us to conclude that the long-sought enhancement of electron coupling to spin fluctuations may be associated with a resonant interaction with extremely shallow electron pockets resulting from dynamic antiferromagnetic ordering. Time-resolved ARPES experiments could shed light on this scenario.

Multiband high-temperature superconductors offer a potential solution to the quantum decoherence problem in superconducting computers. However, the design of superconducting qubits must account for the presence of sign-changing superconducting order parameters.

Iron-based superconductors hold promise for the search for topological super-conductivity. Topological semimetals by carrying exotic electronic state like Weyl and Dirac fermions and having robust surface states open up a lot of applications in electronics and spintronics.

**Acknowledgements**. We are grateful to A. Yaresko and S. Borisenko for discussions. The work was supported by the National Research Foundation of Ukraine (Project 2020.02/0408).

*В.В. Безгуба* [1,2], *О.А. Кордюк* [1,2]

[1] Київський академічний університет,
  бульв. Академіка Вернадського, 36, 03142 Київ, Україна

[2] Інститут металофізики ім. Г.В. Курдюмова НАН України,
  бульв. Академіка Вернадського, 36, 03142 Київ, Україна

БАГАТОЗОННІ КВАНТОВІ МАТЕРІАЛИ

Квантові матеріали визначаються емерджентністю — появою нових властивостей, що є результатом колективних квантових ефектів, а також перспективністю їх для квантових застосувань. Нові надпровідники, від високотемпературних купратів і надпровідників на основі заліза до скручених моношарів, демонструють вищий рівень емерджентної складности, причому багатозонна електронна структура відіграє ключову роль у їх розумінні та потенційному застосуванні. Тут ми робимо короткий огляд ключових багатозонних ефектів у цих надпровідниках і топологічних напівметалах, пропонуючи алґоритми теоретично обґрунтованої розробки нових квантових матеріалів і пристроїв на їхній основі.

**Ключові слова:** надпровідники, топологічні матеріали, електронна фазова діаграма, електронна зонна структура, квантові технології.